\documentclass[a4paper]{jpconf}
\usepackage{graphicx}

\usepackage{txfonts}
\usepackage{bm}
\usepackage{ulem}
\usepackage{braket}
\usepackage{amssymb}
\usepackage{color}

\definecolor{darkgreen}{rgb}{0,0.5,0}

\begin{document}
\title{Simulated floating zone method}

\author{Ryo Ozawa, Yasuyuki Kato, and Yukitoshi Motome}

\address{Department of Applied Physics, University of Tokyo, Tokyo 113-8656, Japan}

\ead{ozawa@aion.t.u-tokyo.ac.jp}

\begin{abstract}
This paper provides the simulated floating zone (SFZ) method that is an efficient simulation technique to obtain 
thermal equilibrium states, especially useful when
domain formation prevents the system from reaching a spatially-uniform stable state. 
In the SFZ method, the system is heated up locally, and the heated region is steadily shifted, 
similar to the floating zone method for growing a single crystal 
with less lattice defect and impurity in experiments. 
We demonstrate that 
the SFZ method suppresses domain formation and accelerates the optimization of the state,
taking an example of a magnetic vortex crystal state realized in itinerant magnets.
We show that the efficiency is maximized
when the local heating temperature is tuned to be comparable to the characteristic energy scale of the ordered state.
\end{abstract}

\section{Introduction}\label{sec:intro}
Numerical simulation is one of the most powerful methods for searching 
thermal equilibrium states of the system.
It has elucidated numerous novel properties that  
one could hardly deal with analytically.
One of the standard techniques is the Markov-chain Monte Carlo sampling on the basis of the Metropolis algorithm, e.g., the single-spin flip update used for classical spin systems. 
A common difficulty in such numerical simulations is that the state is often  
frozen into some configuration and hardly updated, 
especially when the system has a  peculiar energy landscape with multiple local minima or 
almost flat energy dependence in the phase space.
For instance, in some classical spin systems, the spin configuration is 
frozen in a multiple-domain state, and hardly escapes from the metastable state by local spin-flip update.
For avoiding such freezing problems, 
a variety of the simulation techniques have been proposed, 
such as the simulated annealing~\cite{kirkpatrick1984optimization}, 
cluster update~\cite{WANG1990565, PhysRevLett.62.361}, 
and replica exchange methods~\cite{PhysRevLett.57.2607, doi:10.1143/JPSJ.65.1604}. 

In this paper, we propose an efficient technique, 
which will be useful for preventing 
the system from being frozen into multiple-domain states.
Our method is similar to the floating zone method 
for growing a single crystal in experiments. 
Specifically,  a part of the system is heated up to `melt' the frozen structure, and the 
molten region is shifted smoothly to sweep away local defects or domain boundaries during the simulation. 
The procedure helps to grow a single-domain state efficiently, similar to the crystal growth in experiments. 
That is the reason why we
call this method the ``simulated floating zone (SFZ) method".
We demonstrate the efficiency of the SFZ method by applying it to a complicated magnetic ground state 
in the Kondo lattice model with classical localized spins.
We show that the multiple-domain structures are quickly swept away and 
the system reaches a single-domain state.
We find that the efficiency of the SFZ method is optimized 
by setting the local heating temperature comparable to the 
characteristic energy scale of the ordered phase.

The rest of the paper is organized as follows.
In Sec.~\ref{sec:method}, we introduce the fundamental idea of the SFZ method.
In Sec.~\ref{sec:appl}, we demonstrate the efficiency of the SFZ method by applying it to the Langevin dynamics simulation for the Kondo lattice model. 
We summarize our results in Sec.~\ref{sec:summary}.

\section{Method}\label{sec:method}
The idea of the SFZ method is simple. 
It follows the floating zone method used in the single crystal growth. 
In the floating zone method, a narrow region of the crystal is melted by heating, 
and this molten region is shifted along the crystal in a rod shape. 
The procedure suppresses the grains and domains efficiently, and results in a purified single crystal. 
We can introduce the similar technique in numerical simulations for obtaining spatially-uniform equilibrium state. 
Thus, the SFZ method consists of the following steps: 
(1) a part of the system is heated up 
so as to melt the metastable configuration,
(2) the molten region is moved smoothly along the system, and
(3) the procedures are repeated 
a sufficient number of times, until a spatially-uniform stable state is reached.

\begin{figure}[h]
	\begin{center}
	\includegraphics[width=8.0cm]{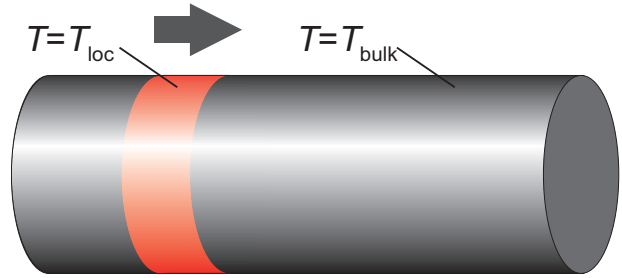}
	\end{center}
	\caption{\label{fig:SFZ_schematic} 
		Schematic picture of the SFZ method. 
		The red region is the heated region, which is smoothly shifted in the direction of the arrow.
	}
\end{figure}

Figure~\ref{fig:SFZ_schematic} shows a schematic picture of the SFZ method. 
The red region is the heated region, 
whose temperature is set at a higher temperature ($T=T_{\rm loc}$) 
than the other parts of the system ($T=T_{\rm bulk}$). 
The heated region is shifted along the system in the direction of the arrow. 
The size and shape of the heated region as well as the speed and direction of its movement are arbitrary: 
one can tune them so as to reach the stable configuration quickly.

The SFZ method is easy to implement and generally  
compatible with many simulation techniques. 
It is also applicable to a wide range of systems, both classical and quantum, in any spatial dimensions.

\section{Application}\label{sec:appl}
In this section, we demonstrate the efficiency of the SFZ method 
in the numerical search of the stable ground state, 
where domain formation makes difficult to reach a spatially-uniform state. 
As a typical example, we here consider a noncoplanar spin texture emergent 
from the interplay between charge and spin degrees of freedom in electrons. 
Specifically, we focus on a 
magnetically-ordered state composed of a periodic arrangement of spin vortices (meron, i.e., half-skyrmion), which we recently found as a candidate for the ground state 
of the Kondo lattice model with classical localized spins on a square lattice~\cite{ozawa2015vortex}.
The Hamiltonian is given by 
\begin{equation}
\mathcal{H} = -\sum_{i,j} \sum_s t_{ij} (\hat{c}^\dagger_{is} \hat{c}^{\;}_{js} + {\rm h.c.})
 - J\sum_{i}\sum_{s, s'} \hat{c}^\dagger_{is}{\bm \sigma}_{ss'}\hat{c}^{\;}_{is'}
\cdot {\bf S}_i,
\label{eq:KLM}
\end{equation}
where $\hat{c}^\dagger_{is}{(\hat{c}^{\;}_{is})}$ is a creation (annihilation) operator of an itinerant electron with spin $s=\uparrow, \downarrow$ at site $i$, ${\bm \sigma}=(\sigma_x, \sigma_y, \sigma_z)$ is the vector representation of the Pauli matrices, and ${\bf S}_i$ is the classical localized spin at site $i$ with $|{\bf S}_i|=1$.
The first term in Eq.~(\ref{eq:KLM}) is the hopping term of itinerant electrons with transfer integral $t_{ij}$: here, we consider only the nearest- and third-neighbor hoppings, $t_1=1$ and $t_3=-0.5$, respectively, on the square lattice. 
The second term in Eq.~(\ref{eq:KLM}) is the onsite Hund's-rule coupling with the coupling constant $J$.
We set the lattice constant $a=1$, the reduced Planck constant $\hbar=1$, and the Boltzmann constant $k_{\rm B}=1$.
 
\begin{figure}
\begin{center}
\includegraphics[width=16.0cm]{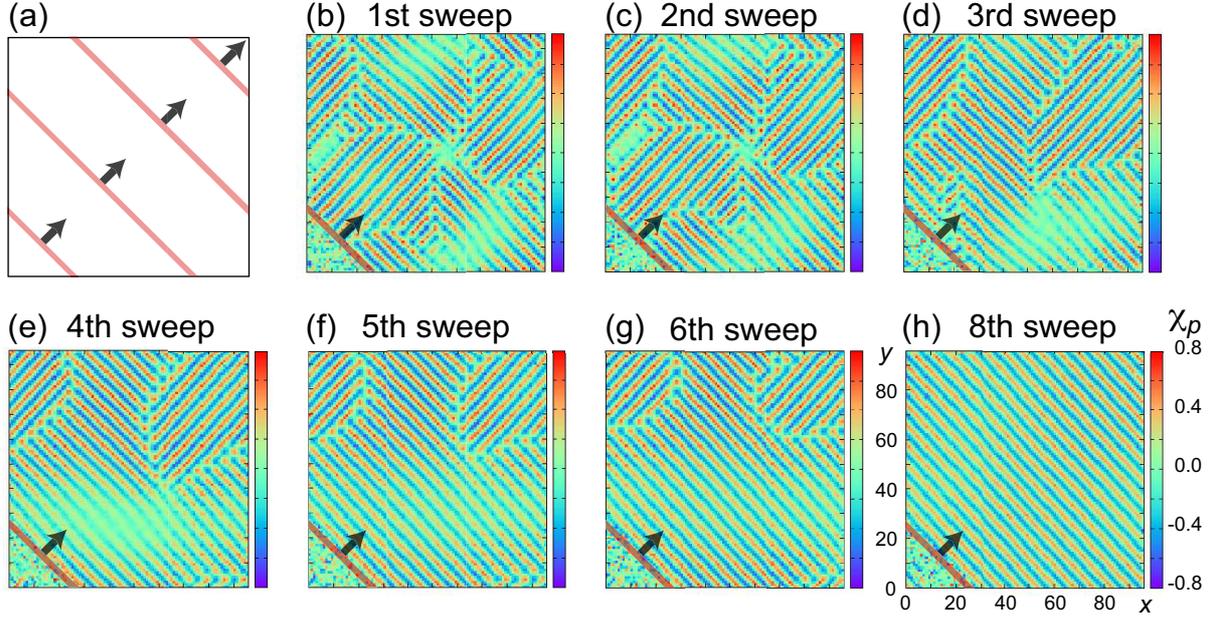}
\end{center}
\caption{\label{fig:D_sweeps} 
(a) Schematic of the SFZ method applied to the model in Eq.~(\ref{eq:KLM}) on the square lattice.
The red region represents the heated strip. 
The arrow indicates the direction of the movement of the strip: 
the strip moves from the bottom-left to top-right corner, 
and restarts from the bottom-left corner in the next cycle.
(b)-(h) Real-space distributions of the spin scalar chirality [Eq.~(\ref{eq:chi})] in the simulation with the use of the SFZ method after the sweeps indicated in each figure. 
The simulation was performed at $t_1=1$, $t_3=-0.5$, and $J=0.05$ on the square lattice with $96^2$ sites.
}
\end{figure}

The vortex crystal state that we found in Ref.~\cite{ozawa2015vortex} exhibits stripes of the spin scalar chirality in the diagonal direction of the square lattice. 
The spin scalar chirality  is defined for each square plaquette $p$ as 
\begin{equation}
\chi_p = \frac{1}{4}\left[\left({\bf S}_{p_1}\times{\bf S}_{p_2}\right)\cdot{\bf S}_{p_3} + \left({\bf S}_{p_4}\times{\bf S}_{p_1}\right)\cdot{\bf S}_{p_2} + \left({\bf S}_{p_3}\times{\bf S}_{p_4}\right)\cdot{\bf S}_{p_1} + \left({\bf S}_{p_2}\times{\bf S}_{p_3}\right)\cdot{\bf S}_{p_4}\right],   
\label{eq:chi}
\end{equation} 
where sites $p_1, \cdots, p_4$ are the vertices of the plaquette $p$ in the counterclockwise direction.
Thus, the vortex crystal state has trivial twofold degeneracy with respect to the directions of chirality stripes, 
which are connected by in-plane fourfold rotation, 
in addition to the global rotation in the spin space. 
It is this twofold degeneracy that leads to domain formation in the current system, as 
discussed later in Fig.~\ref{fig:D_sweeps}(b).

We employ the SFZ method in the numerical simulation using the Langevin dynamics 
for the update of the configurations of localized spins~\cite{Barros_PhysRevB.88.235101}. 
The Langevin dynamics of the localized spins 
is described by the Landau-Lifshitz-Gilbert equation~\cite{gilbert2004phenomenological} as
\begin{equation}
\frac{d{\bf S}_i}{d \tau} = -{\bf S}_i\times{\bf H}_i - \gamma{\bf S}_i\times\left({\bf S}_i\times{\bf H}_i \right),
\label{eq:LLG}
\end{equation}
where the first (second) term describes a precession (dumping) of spins. 
${\bf H}_i$ is the internal effective magnetic field acting on ${\bf S}_i$, which is given by
\begin{equation}
{\bf H}_i = -\frac{\partial \Omega(\{ {\bf S}_j \})}{\partial {\bf S}_i} + {\bf h}_i(\tau, T). 
\label{eq:H_i}
\end{equation}
Here, $\Omega(\{ {\bf S}_i \})$ is the grand potential for a given spin configuration $\{ {\bf S}_i \}$, which is calculated by using the kernel polynomial method~\cite{Weis_RevModPhys.78.275}; ${\bf h}_i$ 
describes thermal fluctuations in the Langevin dynamics, 
which satisfies 
$\langle  h^{\mu}_i(\tau_1, T)  h^{\nu}_j(\tau_2, T)\rangle_\tau =  2T\delta(\tau_1-\tau_2)\delta_{ij}\delta_{\mu\nu}$ %} 
[$\langle \cdots \rangle_\tau$ is the time average 
and $h^\mu_i$ is the $\mu$ component of ${\bf h}_i$ ($\mu=x$, $y$, $z$)]. 
In this thermal-fluctuation term, we introduce the SFZ method. 
The system within a narrow diagonal strip is heated
 up to a temperature $T=T_{\rm loc}$, 
and moved along another diagonal direction; see the schematic picture in Fig.~\ref{fig:D_sweeps}(a); we take the heated strip in the diagonal direction because the chiral stripes appear along the diagonal (or $\pi/4$ rotated) direction, as shown below.

Figure~\ref{fig:D_sweeps} shows the results of the Langevin dynamics simulation 
accelerated by the SFZ method.
We consider the model in Eq.~(\ref{eq:KLM}) with $J = 0.05$ on the square lattice 
with $96^2$ sites and periodic boundary conditions.  
We set the electron filling so as to realize 
the vortex crystal with ordering vectors $ \sim (\pm\pi/4, \pi/4)$~\cite{ozawa2015vortex}. 
The Langevin dynamics simulation is performed by taking $\gamma=1$ and the time interval as $\Delta \tau=30$.
In the kernel polynomial method,  we take the Chebyshev order $M=800$ and 
the number of random vectors $R=144$. 
We use the heated strip with four-site width 
(we do not impose the periodic boundary conditions on the strip shape), and set the temperature at $T_{\rm loc} = 0.004$, 
while taking $T_{\rm bulk}=0$. 
The heated strip is shifted in a constant speed along the system 
so that one sweep from the bottom-left to top-right corner takes a time of $4.8\times 10^4$: the Langevin dynamics update is performed for $1.6\times 10^3$ times during the single sweep. 
The next sweep starts from the same bottom-left corner just after the previous sweep ends. 

Figure~\ref{fig:D_sweeps}(b) shows a typical multiple-domain state obtained 
by the Langevin dynamics simulation without the SFZ method staring from a random spin configuration.
We here plot the real-space distribution of the scalar chirality $\chi_p$ defined in Eq.~(\ref{eq:chi}). 
The system has several domains; each domain has chiral stripes running in the diagonal-upward or downward direction.
In this state, the spin configuration is almost frozen: 
the domain walls are hardly removed 
by the update in the Langevin dynamics. 
Once we turn on the SFZ method, however, the multiple domains are smeared and the system reaches a single-domain state.
Figures~\ref{fig:D_sweeps}(c)-\ref{fig:D_sweeps}(h) show the evolution of the real-space distributions of $\chi_p$ 
during the simulation with the use of the SFZ method.
As shown in the figures, when we repeat the sweep, one type of the two degenerated 
domains grows and merges into larger domains.  
After eight cycles, we successfully obtain the spatially-uniform state, 
as shown in Fig.~\ref{fig:D_sweeps}(h).

We note that the direction of the chiral stripes in the final uniform state is always parallel to the heated strip: 
when we rotate the strip and its motion by $\pi/2$ in the SFZ simulation, 
the other domain with $\pi/2$-rotated stripes 
dominates the system in the final uniform state.
Such a selection of energetically-degenerate domains 
might occur in the SFZ method, 
depending on the microscopic energetics in the system as well as 
the details of the floating zone. 
In the current case, we note that the selected domain has 
the dominant spin helix parallel to the heated strip~\cite{ozawa2015vortex}.  
The reason why this domain is chosen is presumably because the disturbance of electron motion along the spin helix leads to a substantial energy loss in the spin state.
 
\begin{figure}
\begin{center}
\includegraphics[width=16.0cm]{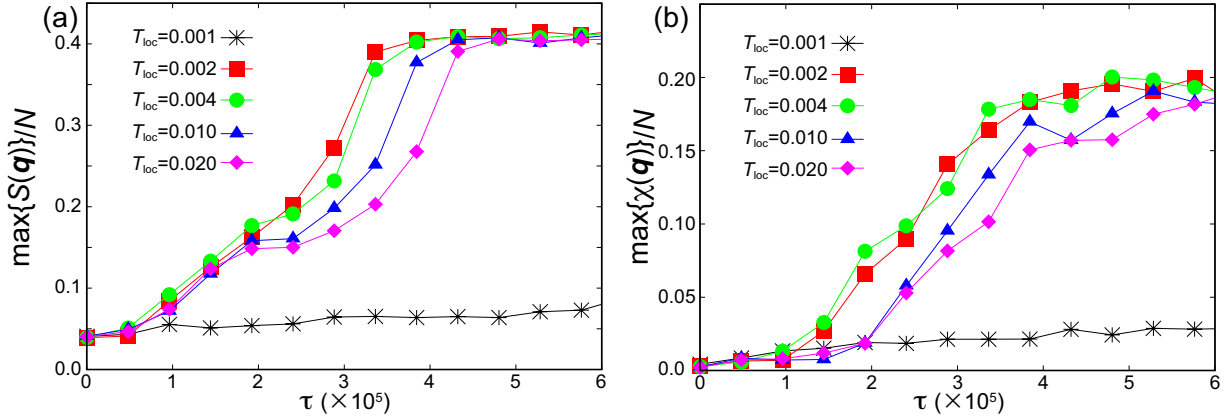}
\end{center}
\caption{\label{fig:sweep_Ts} 
Time evolution of the largest peak (a) in the spin structure factor, $S({\bf q})$, and (b) in the chirality structure factor, $\chi({\bf q})$, divided by the system size for different temperatures of the heated region in the SFZ method, $T_{\rm loc} = 0.001$, $0.002$, $0.004$, $0.010$, and $ 0.020$.
} 
\end{figure}

Let us discuss the efficiency of the SFZ method 
while changing the heating temperature $T_{\rm loc}$. 
In order to measure the efficiency, 
we monitor the time evolution of the dominant peaks of the structure factors 
for spin and chirality degrees of freedom. 
The structure factors are defined by
\begin{eqnarray}
S({\bf q})    = \frac{1}{N} \sum_{i,j} {\bf S}_i\cdot {\bf S}_je^{i{\bf q}\cdot({\bf r}_i-{\bf r}_j)},\quad 
\chi({\bf q}) = \frac{1}{N} \sum_{p,q} \chi_p\chi_qe^{i{\bf q}\cdot({\bf r}_{p_1}-{\bf r}_{q_1})},
\end{eqnarray}
where $N$ is the system size and ${\bf r}_i$ is the real-space coordinate of site $i$. 

Figure~\ref{fig:sweep_Ts} shows the time evolution of the largest peaks of 
$S({\bf q})$ and $\chi({\bf q})$ in the SFZ simulation for several different $T_{\rm loc}$
by starting from the same initial state shown in Fig.~\ref{fig:D_sweeps}(b).
Here, we plot the data at every SFZ sweep. 
When we set $T_{\rm loc}=0.001$,  
both peaks grow very slowly and do not  change substantially from the initial values after the long simulation time. 
This is presumably 
because $T_{\rm loc}=0.001$ is too low to melt the frozen domain structure. 
For $T_{\rm loc}\geq 0.002$, both peaks grow during the simulation, and finally saturate at the same values, 
which are expected for the single-domain state, after the sufficiently long simulation time.
When further increasing $T_{\rm loc}$, however, the time  
necessary for the convergence becomes longer. 
For both $S({\bf q})$ and $\chi({\bf q})$, 
$T_{\rm loc}\simeq0.002$-$0.004$ 
optimizes the efficiency of the SFZ simulation in the current case.

We note that the optimal value of $T_{\rm loc}$ is comparable to 
the energy scale of the magnetic ordering. 
In the Kondo lattice model in the small $J$ region, the energy scale 
is set by the so-called Ruderman-Kittel-Kasuya-Yosida (RKKY) interaction, which is given by the second-order perturbation in terms of $J$~\cite{Ruderman, Kasuya, Yosida1957}.
In the current case, the energy scale is about $J^2 = 0.0025$, 
which is close to the optimal $T_{\rm loc} \simeq 0.002$-$0.004$. 
Thus, the result suggests that the SFZ method works most efficiently 
when the heating temperature is comparable to the melting temperature 
of the corresponding ordered phase. 
This is presumably because the small part of the system included in the heated strip is easily aligned in a spatially-uniform ordered state owing to the growing correlation length near the melting temperature.

\section{Summary}\label{sec:summary}
In summary, we have introduced the SFZ method as an efficient technique to obtain the thermal equilibrium state 
in numerical simulations. 
In this method, a part of the system is locally heated up and shifted smoothly 
all over the system. 
The main purpose of using this method is 
to accelerate the numerical convergence, 
preventing the state from being 
frozen in a local minimum or a flat energy landscape in the phase space.
The advantage is that this method is simple and applicable to many situations.
In this paper,
we have applied the SFZ method to the Langevin dynamics simulations for the square Kondo lattice model 
as an example
where domain formation is often problematic for 
obtaining the ground state in the simulations without the SFZ method.
We have demonstrated that the introduction of
the SFZ method sweeps away the domain structures and substantially accelerates the convergence to the spatially-uniform state. 
In addition, we have found that the convergence is 
most accelerated by setting the local heating temperature comparable to the characteristic energy scale of the ordered state.

\section{Acknowledgments}
%Authors wishing to acknowledge K. Barros for the fruitful discussions on the KPM-LD simulations. 
The SFZ methods based on the Langevin dynamics simulations were carried out at the Supercomputer Center, 
Institute for Solid State Physics, University of Tokyo.
R.O. is supported by the Japan Society for the Promotion of Science 
through a research fellowship for young scientists and the Program for Leading Graduate Schools (ALPS). 
This research was supported by KAKENHI (No.~24340076, 26800199), the Strategic Programs for Innovative Research (SPIRE), MEXT, and the Computational Materials Science Initiative (CMSI), Japan. 
\section*{References}
\bibliographystyle{iopart-num}
%\bibliography{reference}

\end{document}